\documentclass[12pt]{article}%
\usepackage{amsmath,latexsym}
\usepackage{graphicx}
\usepackage{amsmath}
\usepackage{amsfonts}
\usepackage{amssymb}%
\begin{document}

\title{{On the nature of exotic matter in
   Morris-Thorne wormholes}}
   \author{
Peter K F Kuhfittig*\\  \footnote{kuhfitti@msoe.edu}
 \small Department of Mathematics, Milwaukee School of
Engineering,\\
\small Milwaukee, Wisconsin 53202-3109, USA}

\date{}
 \maketitle

\begin{abstract}\noindent
It is well known that phantom energy, which is
characterized by the equation of state
$p=\omega\rho$, $\omega <-1$, can support
Morris-Thorne wormholes since it leads to a
violation of the null energy condition.  The
purpose of this note is to show that the
converse is also true in the following sense:
for a typical shape function, the equation
of state of exotic matter in the vicinity of
the throat is given by $p_r=\omega\rho$,
$\omega <-1$, where $p_r$ is the radial
pressure.  Some implications thereof are
also considered.   

\end{abstract}

\section{Introduction}\label{S:introduction}

Wormholes are tunnel-like structures in
spacetime that connect widely separated
regions of our Universe or different universes
in a multiverse.  Apart from some forerunners,
macroscopic traversable wormholes were first
studied in detail by  Morris and Thorne
\cite{MT88} in 1988.  They had proposed the
following static and spherically symmetric
line element for a wormhole spacetime:
\begin{equation}\label{E:line}
ds^{2}=-e^{2\Phi(r)}dt^{2}+\frac{dr^2}{1-b(r)/r}
+r^{2}(d\theta^{2}+\text{sin}^{2}\theta\,
d\phi^{2}),
\end{equation}
using units in which $c=G=1$.  Here
$\Phi=\Phi(r)$ is called the \emph{redshift
function}, which must be everywhere finite
to prevent an event horizon.  The function
$b=b(r)$ is called the \emph{shape function}
since it determines the spatial shape of the
wormhole when viewed, for example, in an
embedding diagram \cite{MT88}.  The spherical
surface $r=r_0$ is called the \emph{throat}
of the wormhole, where $b(r_0)=r_0$.  The
shape function must also meet the requirement
$b'(r_0)<1$, called the \emph{flare-out
condition}, while $b(r)<r$ for $r>r_0$.  We
also require that $b'(r_0)>0.$

The flare-out condition can only be met by
violating the null energy condition (NEC)
which states that
\begin{equation}
  T_{\alpha\beta}k^{\alpha}k^{\beta}\ge 0
\end{equation}
for all null vectors $k^{\alpha}$, where
$T_{\alpha\beta}$ is the energy-momentum
tensor.  Matter that violates the NEC is
called ``exotic" in Ref. \cite{MT88}.
Exotic matter is usually confined to a
narrow region around the throat.  In
particular, for the outgoing null vector
$(1,1,0,0)$ the violation of the NEC takes 
on the form
\begin{equation}
   T_{\alpha\beta}k^{\alpha}k^{\beta}=
   \rho +p_r<0,
\end{equation} 
Here $T^t_{\phantom{tt}t}=-\rho$ is the energy
density, $T^r_{\phantom{rr}r}= p_r$ is the
radial pressure, and
$T^\theta_{\phantom{\theta\theta}\theta}=
T^\phi_{\phantom{\phi\phi}\phi}=p_t$ is
the lateral (transverse) pressure.  Before
continuing, let us list the Einstein field
equations:

\begin{equation}\label{E:Einstein1}
  \rho(r)=\frac{b'}{8\pi r^2},
\end{equation}
\begin{equation}\label{E:Einstein2}
   p_r(r)=\frac{1}{8\pi}\left[-\frac{b}{r^3}+
   2\left(1-\frac{b}{r}\right)\frac{\Phi'}{r}
   \right],
\end{equation}
\begin{equation}\label{E:Einstein3}
   p_t(r)=\frac{1}{8\pi}\left(1-\frac{b}{r}\right)
   \left[\Phi''-\frac{b'r-b}{2r(r-b)}\Phi'
   +(\Phi')^2+\frac{\Phi'}{r}-
   \frac{b'r-b}{2r^2(r-b)}\right].
\end{equation}
Since Eq. (\ref{E:Einstein3}) can be obtained
from the conservation of the stress-energy tensor
$T^{\mu\nu}_{\phantom{\mu\nu};\nu}=0$, only
Eqs. (\ref{E:Einstein1}) and (\ref{E:Einstein2})
are actually needed.

The discussion of exotic matter took a
dramatic turn with the discovery that the
Universe is expanding at an accelerated  rate
caused by a hypothesized \emph{dark energy}
\cite{aR98, PTW99}, which makes up about
70\% of the substance  in the Universe.
Dark energy is characterized by its
barotropic equation of state $p=
\omega\rho$, where $\rho >0$ and
$\omega <-1/3$.  The range $\omega <-1$,
called \emph{phantom energy}, is of
particular interest to us since it leads
to a violation\emph{} of the
 NEC \cite{sS05, fL05}:
\begin{equation}
   p+\rho=\omega\rho +\rho=(\omega +1)
      \rho   <0.
\end{equation}
To apply the equation of state $p=\omega\rho$
to wormholes, where the pressure is generally
anisotropic, one regards the pressure in
the equation of state as a negative radial
pressure, and the tangential pressure is
obtained from the Einstein field equations.
Moreover, according to Ref. \cite{SK04},
moving away from the throat, the radial
and transverse pressures rapidly become
equal, showing that phantom energy can in
principle support traversable wormholes.

The main purpose of this note is to
obtain a converse in the following sense:
for any Morris-Thorne wormhole with a
typical shape function, the exotic matter
near the throat has the equation of state
$p_r=\omega\rho$, $\omega <-1$  (Sections
\ref{S:introduction}-\ref{S:state}).
Some implications of these ideas are
discussed in Sections \ref{S:phantom}
and \ref{S:radial}.


\section{A generic shape function}
     \label{S:shape}

Consider the family of functions \cite{MT88}
\begin{equation}\label{E:shape}
   b_{\eta}(r)=r_0\left(\frac{r}{r_0}
      \right)^{1-\eta},\quad 0<\eta<1.
\end{equation}
Evidently, $b_{\eta}(r_0)=r_0$, while
\begin{equation}\label{E:bprime}
    0<b'_{\eta}(r_0)=1-\eta<1,
\end{equation}
thereby meeting the basic requirements of
a shape function.  The special case
$\eta=1/2$ leads to the parabola $b(r)
=\sqrt{r_0r}$, which is concave down to
the right of $r=r_0$ with $b(r)<r$ for
all $r$. The same behavior is exhibited
by all members of the family in Eq.
(\ref{E:shape}).  Moreover, for every
value of $m$ between 0 and 1, there
exists a member $b_{\eta}(r)$ such that
$b'_{\eta}(r_0)=m$.

These general properties are enough to
suggest that any shape function of a
Morris-Thorne wormhole can be closely
approximated by a member of
$\{b_{\eta}(r)\}$, at least in the
vicinity of the throat.  To that end,
we need to assume that a typical shape
function $b=b(r)$ is concave down in
the immediate vicinity of $r=r_0$ with
$b(r)<r$.  In other words, $b(r)$ has
the following properties:
\\
\phantom{rrrrr} I. $b(r_0)=r_0$,
where $r=r_0$ is the throat of the wormhole,
\\
\phantom{rrrrr}II. $0<b'(r_0)<1$,
\\
\phantom{vvv}III. $b(r)$ is concave down
for $r$ near $r=r_0$.
\\Property I is automatically satisfied
by $b_{\eta}(r)$.  Regarding Property II,
the value of $b'(r_0)$ determines $\eta$ by
Eq. (\ref{E:bprime}).  Finally, for any
$\eta\,\,\varepsilon\, (0,1)$,
\[
    b''(r_0)=\frac{1}{r_0}(1-\eta)(-\eta)
    <0,
\]
so that $b_{\eta}(r)$ is indeed concave
down near $r=r_0$ (Property III).  As
a result, any shape function that meets
Conditions I, II, and III can be
approximated by some $b_{\eta}(r)$ in
the vicinity of the throat.

\section{The equation of state}
   \label{S:state}
By making use of the properties of $b=b(r)$
discussed in Sec. \ref{S:shape}, we are
going to obtain the equation of state for
exotic matter in the vicinity of the
throat.  As already noted, this is
precisely the region where we would
expect to find exotic matter.  First we
observe that in Eq. (\ref{E:Einstein2}),
the second term inside the brackets can
be neglected.  We now obtain
\begin{equation}\label{E:density}
   \rho(r)=\frac{1}{8\pi}\frac{1}{r^2}
   (1-\eta)\left(\frac{r_0}{r}\right)
   ^{\eta}
\end{equation}
and
\begin{equation}\label{E:pressure}
   p_r(r)=\frac{1}{8\pi}\left(-\frac{1}{r^3}
   \right)(r_0)\left(\frac{r}{r_0}
   \right)^{1-\eta}.
\end{equation}
Eqs. (\ref{E:density}) and (\ref{E:pressure})
lead to the surprising result that
$\rho(r)/p_r(r)$ reduces to a constant value
for all $r$:
\begin{equation}\label{E:rad}
   \frac{\rho(r)}{p_r(r)}=\eta-1\quad
   \text{or}\quad p_r(r)=
   \frac{1}{\eta-1}\rho(r).
\end{equation}
If we now write
\begin{equation}\label{E:EoS}
   p_r(r)=\omega\rho(r),
\end{equation}
we conclude that, since $0<\eta<1$,
\[
    \omega<-1.
 \]
So Eq. (\ref{E:EoS}) has the same form as
the equation of state for phantom energy,
to be discussed further in the next section.
Our main conclusion is that the unavoidable
exotic matter not only violates the null
energy condition, it can be characterized
by the equation of state (\ref{E:EoS}),
where $\omega <-1$.

\section{Returning to phantom energy}
      \label{S:phantom}
While Eq. (\ref{E:EoS}) has the same form as
the cosmological equation of state
$p(r)=\omega\rho(r)$ for phantom energy, there
is one critical difference.  For the latter,
both $p$ and $\rho$ are necessarily small.
For the former, $p_r(r)=\omega\rho(r)$, the
opposite is true, as already pointed out
in Ref. \cite{MT88}: recalling that the
tension $\tau(r)$ is the negative of $p(r)$,
we can rewrite Eq. (\ref{E:Einstein2}) as
(reintroducing $c$ and $G$ for now)
\begin{equation}\label{E:tau}
   \tau(r)=\frac{b(r)/r-2[r-b(r)]\Phi'(r)}
   {8\pi Gc^{-4}r^2}.
\end{equation}
According to Ref. \cite{MT88}, the radial
tension at the throat is
\begin{equation}\label{E:tension}
   \tau=\frac{1}{8\pi Gc^{-4}r_0^2}\approx
   5\times 10^{41}\frac{\text{dyn}}{\text{cm}^2}
   \left(\frac{10\,\text{m}}{r_0}\right)^2.
\end{equation}
In particular, for $r_0=3\,\,\text{km}$,
$\tau$ has the same magnitude as the pressure
at the center of a massive neutron star.  For
a phantom-energy wormhole, such an outcome is
clearly absurd: for such a wormhole to exist,
$r=r_0$ would have to be extremely large in
 view of Eq. (\ref{E:tension}).  So
 phantom-energy wormholes could only exist
 on very large scales.

\section{The large radial tension revisited}
   \label{S:radial}
It is proposed in Ref. \cite{pK20} that the
enormous radial tension can be attributed to
the effect of noncommutative geometry.  This
conclusion actually follows quite readily from
Eq. (\ref{E:rad}).  First we need to recall that
in noncommutative geometry, an offshoot of string
theory, coordinates may become noncommuting
operators on a $D$-brane \cite {eW96, SW99}.
This statement refers to the commutator
$[\textbf{x}^{\mu},\textbf{x}^{\nu}]
=i\theta^{\mu\nu}$, where $\theta^{\mu\nu}$
is an antisymmetric matrix.  Moreover,
noncommutativity replaces point-like
structures by smeared objects, thereby
elimination the divergences that normally
occur in general relativity \cite{NSS06,
NS10}.  A natural way to accomplish
the smearing effect is to assume that the
energy density of a static and spherically
symmetric and particle-like gravitational
source has the form \cite{NM08, pK15}
\begin{equation}\label{E:rho}
  \rho(r)=\frac{\mu\sqrt{\beta}}
     {\pi^2(r^2+\beta)^2}.
\end{equation}
This form can be interpreted to mean that
the mass $\mu$ of the particle is diffused
throughout the region of linear dimension
$\sqrt{\beta}$ due to the uncertainty.  An
important observation in noncommutative
geometry is that whenever we make use of
Eq. (\ref{E:rho}), we can keep the standard
forms of the Einstein field equations since
the Einstein tensor retains its original
form; only the stress-energy tensor is
modified \cite{NSS06}.  So the length
scales can be macroscopic.  The shape
function $b(r)$ can now be determined
from Eqs. (\ref{E:rho}) and (\ref
{E:Einstein1}):
\begin{equation}\label{E:shapefunction}
   b(r)=\frac{4M\sqrt{\beta}}{\pi}
  \left(\frac{1}{\sqrt{\beta}}\text{tan}^{-1}
  \frac{r}{\sqrt{\beta}}-\frac{r}{r^2+\beta}-
  \frac{1}{\sqrt{\beta}}\text{tan}^{-1}
  \frac{r_0}{\sqrt{\beta}}+\frac{r_0}{r_0^2
  +\beta}\right)+r_0,
\end{equation}
where $M$ is the mass of the wormhole.
At this point, all we need to observe is
that $b(r)$ in Eq. (\ref{E:shapefunction})
satisfies Conditions I, II, and III in
Section \ref{S:shape}.  So by Eq.
(\ref{E:rad}), the radial pressure has
the form
\begin{equation}\label{E:rad2}
   p_r(r)=\frac{1}{\eta -1}
   \frac{\mu\sqrt{\beta}}
   {\pi^2(r^2+\beta)^2}.
\end{equation}
To  apply this equation to wormholes, we
need to return to Ref. \cite{pK20}.  One
can assume that the throat surface $r=r_0$
consists of a set of smeared particles,
resulting in a smeared surface.  According
to the discussion in Ref. \cite{pK20}, the
energy density of the surface is given by
\begin{equation}
   \rho_s=\frac{m\sqrt{\beta}}
   {\pi^2[(r-r_0)^2+\beta]^2},
\end{equation}
where $m$ is the mass of the surface.
Eq. (\ref{E:rad2}) can now be replaced by
\begin{equation}
   p_r(r)=\frac{1}{\eta -1}
   \frac{m\sqrt{\beta}}
   {\pi^2[(r-r_0)^2+\beta]^2}.
\end{equation}
Since the tension $\tau$ is the negative of
the pressure, we have at or near the throat
$r=r_0$
\begin{equation}
  \tau(r)=\frac{1}{1-\eta}
  \frac{m}{\pi^2(\sqrt{\beta})^3},
\end{equation}
which can be made as large as necessary for \
proper choices of $\eta$ and $\beta$.

\section{Conclusion}

The unavoidable exotic matter in a Morris-Thorne
wormhole is assumed to be confined to the
immediate vicinity of the throat.  In this
note we also assume that a typical shape
function is concave down near $r=r_0$ (in
addition to its usual properties).  It is
subsequently shown that the equation of state
of the exotic matter is $p_r=\omega\rho$,
$\omega<-1$, which is essentially the same
as the equation of state for phantom dark
energy, since, in a phantom-energy wormhole,
the radial and lateral pressures tend to
quickly become equal away from the throat
\cite{SK04}.  So while phantom energy
behaves like exotic matter, the converse
is also true since the equation of state
of exotic matter has the same form.  There
is one essential difference, however: since
$p$ and $\rho$ are necessarily small in the
phantom-energy case, it follows from Eq.
(\ref{E:tension}) that such wormholes could
only exist on very large scales.  It is
shown in Section \ref{S:radial}, however,
that for smaller wormholes, the enormous
radial tension can be attributed to the
effect of a noncommutative-geometry
background, confirming the result in
Ref. \cite{pK20}.

\end{document}